\documentstyle[amssymb,preprint,aps]{revtex}

\begin{document}
\draft
\title{Near-field spectroscopy of a gated electron gas: a direct evidence for
electrons localization}
\author{G. Eytan, Y. Yayon, M. Rappaport, H. Shtrikman, and I. Bar-Joseph}
\address{Department of Condensed Matter Physics,\\
The Weizmann Institute of Science,\\
Rehovot 76100, Israel}
\maketitle

\begin{abstract}
The near-field photoluminescence of a gated two-dimensional electron gas is
measured. We use the negatively charged exciton, formed by binding of an
electron to a photo-excited electron-hole pair, as an indicator for the
local presence of charge. Large spatial fluctuations in the luminescence
intensity of the negatively charged exciton are observed. These fluctuations
are shown to be due to electrons localized in the random potential of the
remote ionized donors. We use these fluctuations to image the electrons and
donors distribution in the plane.
\end{abstract}


\narrowtext

The two-dimensional electron gas (2DEG) which is formed in semiconductor
heterostructures has been a subject of intense research in the last two
decades. The electrons in these heterostructures are provided by a remote
layer of donors and are confined in a two-dimensional potential. Applying a
gate voltage enables one to affect the conductivity of the 2DEG in a
profound way: at low temperatures a drop of several orders of magnitude can
be observed over a small range of gate voltage.\cite{Tsui} This large change
in conductivity is accompanied by a relatively small change in the electron
density, typically less than an order of magnitude. The microscopic
understanding of this behavior is the focus of this paper.

It has been realized that the remote ionized donors, which provide the
electrons to the 2DEG, play an important role in this drop of conductivity 
\cite{Tsui},\cite{Efros} Spatial variations in the density of these donors
are manifested as random potential fluctuations in the plane of the 2DEG. At
small negative gate voltages, when the conductivity is high, these
fluctuations are effectively screened by the 2DEG. As the electron density
is decreased at larger negative gate voltages, the screening becomes less
effective, and the fluctuations grow, giving rise to a large drop in the
electron conductivity.

The photoluminescence (PL) of a gated 2DEG was intensively studied in recent
years.\cite{Gleb},\cite{Shields} It was found that at small negative gate
voltage the PL spectrum is a broad line, with a width of approximately the
Fermi energy (Fig. 1). As the gate voltage becomes more negative and the
electron density is reduced, the PL spectrum narrows. The important finding
was that at a certain gate voltage the spectrum changes abruptly into two
exciton peaks. Decreasing the gate voltage further changes the relative
strength of the two peaks. It was shown\cite{Gleb},\cite{Shields} that the
high energy peak is due to a recombination of a neutral exciton (X), a bound
complex of an electron and a hole. The low energy peak was found to be
associated with the negatively charged exciton (X$^{-}$), which consists of
two electrons and a hole and is a semiconductor analogue of the hydrogen ion
H$^{-}$.\cite{Stebe},\cite{Khang} The appearance of the excitonic peaks is
an indication that the screening of the electron-hole interaction by the
2DEG is suppressed and a bound complex can be formed.\cite{Chemla} Indeed, a
correlation was found between the growth of the X peak and the drop of
conductivity.\cite{Gleb}

It has been shown that the X$^{-}$ is formed in a gated 2DEG by binding of a
photo-excited electron-hole pair to an electron from the 2DEG.\cite{Gleb}, 
\cite{Shields} The X$^{-}$\ can therefore be used as{\it \ a probe for the
presence of electrons, }and by mapping its intensity across the area of the
sample, one can identify the regions where electrons are located.

In this work we use the X$^{-}$ luminescence to image the distribution of
electrons in the 2DEG in the range of gate voltages where the conductivity
drops. We show that in this regime the electrons are localized in the
potential fluctuations of the remote ionized donors. This is done using
near-field spectroscopy, which provides sub-wavelength spatial resolution. 
\cite{Betzig},\cite{NF book} The underlying idea is rather simple: the
spectral signature of a region filled with electrons is very different from
that of an empty region. Thus, by collecting the PL through a small aperture
which is scanned in close proximity to the surface one can map the charge
distribution in the plane.

We have built a low temperature near-field scanning optical microscope
(NSOM), which collects the emitted PL through a tapered optical fiber tip.
The tip, which is manufactured by Nanonics, is coated at the sides by
aluminium and has a clear aperture of $\thicksim 250$ nm and a transmission
of $10^{-3}$. We have verified the continuity of the metallic coating using
SEM imaging. The tip is glued to a commercial tuning fork, having a
resonance frequency of 32,768 Hz. The piezoelectric signal from the tuning
fork is used to control the height of the tip above the sample surface at $%
\sim 10$ nm and to measure the sample topography.\cite{Karrai} The sample is
mounted on a 2 inch piezoelectric tube, giving a scan range of 11x11 $\mu $m$%
^{2}$ at 4 K. A coarse X-Y movement assembly enables movement to different
regions of the sample within an area of 2x2 mm$^{2}$, with 1 $\mu $m steps.
The microscope is inserted into a sealed tube filled with He exchange gas,
and the tube is immersed in a storage dewar. Further details on the NSOM are
given elsewhere.\cite{RSI}

The photo-excitation is done using a second single mode fiber which is
oriented such that the light comes out nearly parallel to the surface,
illuminating thus a broad region of $\thicksim 1$ mm$^{2}$. This mode of
operation, which does not use the tip for illumination,\cite{Hess},\cite
{Gershoni} ensures uniform excitation of the scanned region while
maintaining a good spatial resolution. The excitation intensity is typically
50 mW/cm$^{2}$ at a wavelength of 632.8 nm. The collected PL is dispersed in
a 0.5 m spectrometer and detected in a thermoelectrically cooled,
back-illuminated CCD detector. The system spectral resolution is 0.04 nm.
The typical signal at that excitation condition is 30 - 40 counts/sec at the
exciton peak. Scanning is performed with a typical integration time of 5
seconds per point and a step size of 100 nm. These scanning parameters
translate to a measurement time of several hours for a two dimensional scan.
We have verified that the system is stable during this measurement time. To
determine the spatial resolution of the system we deposited an array of
opaque gold lines (1.5 $\mu $m width and 3.5 $\mu $m period) on a sample of
GaAs quantum well, and measured the collected PL as the tip is scanned
across the pattern. The spatial resolution is obtained by the distance over
which the integrated PL intensity changes from 0.1 to 0.9 of its maximum
value. The value which is obtained is $\sim 250$ nm, in a very good
agreement with the results of the SEM imaging, ensuring there are no leaks
in the matallic coating of the tip.

The sample consists of a 20 nm GaAs quantum well followed by 37.5 nm Al$%
_{0.37}$Ga$_{0.63}$As spacer layer and a 10 nm Al$_{0.37}$Ga$_{0.63}$As
layer doped with silicon at concentration of 3.5x10$^{18}$ cm$^{-3}$. The
structure is capped by a 20 nm undoped Al$_{0.37}$Ga$_{0.63}$As and 10 nm
GaAs. A 2x2 mm$^{2}$ mesa was etched, and ohmic contacts were alloyed into
the 2DEG layer. A 4 nm Pd/Au semi-transparent gate was evaporated on top. We
have verified the continuity and uniformity of the gate by SEM imaging. The
2DEG concentration was 4x10$^{11}$cm$^{-2}$, and the mobility 1.3x10$^{6}$
cm $^{2}$/V sec, both measured at 4.2 K. The laser energy is below the
AlGaAs bandgap, hence no photoexcited carriers are created in the high
bandgap region. Nevertheless, upon excitation with the laser the electron
density is reduced, by an amount which depends on the laser intensity.
Consequently, the gate voltage needed to deplete the sample is intensity
dependent.

Let us now turn to the experimental results. Figure 2A shows four near-field
spectra measured at different locations at the same gate voltage, V$_{\text{%
g }}=-$ $0.135$ V. At this gate voltage the spectrum is excitonic: the high
and low energy peaks are the X and X$^{-}$, respectively. It can be seen
that the near-field spectrum is different from one location to another.
While the height of the X peak is very similar in the four spectra, the
height of the X$^{-}$ peak varies substantially. Since the X$^{-}$ is an
indicator for the presence of electrons, these variations in its intensity
show that the {\it electrons are non-uniformly distributed}.

These PL intensity variations disappear at the far-field, when the PL is
collected far from the surface. To substantiate the fact that the far-field
spectrum is a superposition of different local spectra, we performed the
following procedure. We first scanned a square of 1x1 $\mu $m$^{2}$ and
summed up the PL spectra from all the points (Fig. 2b solid line). We then
withdrew the tip to a distance of 1 $\mu $m from the sample and measured the
PL above the center of this square (Fig. 2B dashed line). It can be seen
that the two curves nearly coincide.

To quantify the fluctuations of the X$^{-}$ intensity we measured the PL
along 11 $\mu $m-long line, with 100 nm steps between subsequent
measurements, and integrated the intensity under the X$^{-}$ peak at each
point. For comparison, we performed the same procedure for the X peak. The
results are shown in Fig. 3A (the four points at which the spectra of Fig.
2A were taken are denoted on the figure). It is evident that there are large
fluctuations in the X$^{-}$ intensity. These fluctuations occur throughout
the scanned region and on any length scale, down to the resolution limit.
The X peak exhibits much smaller fluctuations.It should be emphasized that
these fluctuations are stable over time: the line scan is reproduced by
repeating the measurement over and over again. We found no correlation
between these fluctuations of the X$^{-}$ intensity and the shear-force
signal, which exhibits a $\pm 1.5$ nm noise.

Figure 3B shows a comparison between the integrated intensity under the X$%
^{-}$ peak for three line scans (over the same line as in Fig. 3A), each
scan done at a different gate voltage: V$_{\text{g}}=-0.105$ V, $-0.135$ V
and $-0.155$ V. We normalize the curves by dividing the value at each point
along the line by the average value of that curve. Comparing the different
line scans, one can clearly see that the relative fluctuations amplitude
increases with the gate voltage. The positions of the maxima and minima,
however, are fixed in space, and are almost unaffected by the gate voltage.
Figure 4 summarizes the dependence of the fluctuations amplitude on gate
voltage (note that this measurement is taken at a smaller illumination
intensity then the measurements of Fig. 3, hence the gate voltages are more
negative). The graph describes the dependence of the normalized standard
deviation of the X$^{-}$ fluctuations amplitude on gate voltage. It can be
clearly seen that the fluctuations amplitude is constant over a large range
of gate voltages and starts to increase at the voltage at which the spectrum
becomes excitonic. This correlation between the appearance of the excitonic
spectrum and the rise of the fluctuations amplitude is very significant. The
excitons appearance is an indication for a change in the screening
properties of the 2DEG: it becomes ineffective in screening the
electron-hole interaction. It follows that the 2DEG is also ineffective in
screening the random potential induce by the ionized donors in the doped
AlGaAs layer, and the fluctuations in this potential grow. (A similar
process explains the appearance of D$^{-\text{ }}$centers in lightly-doped
GaAs quantum wells.\cite{Dminus}) These donors are randomly distributed,
with an average distance between them of a few nm. They give rise to a
random electrostatic potential in the 2DEG plane, with the smallest spatial
period being the spacer width (37.5 nm in our sample). \cite{Efros} The high
spatial frequencies of the observed fluctuations and the fixed positions of
the minima and maxima are consistent with this explanation.

It is evident from Fig. 4 that fluctuations in the PL intensity are observed
throughout the gate voltage range. At the range where the PL is 2DEG-like,
namely - non-excitonic and shifted to lower energies, the electrons are
free. The weak fluctuations in this range are independent of the gate
voltage, and are due to localization of the photo-excited holes in the
donors potential. The observed increase in the fluctuations amplitude as the
spectrum becomes excitonic is {\it the onset of the electrons localization}.

Since our spatial resolution is 250 nm, we are unable to collect light from
a single localization site which has a typical size of $\thicksim 40$ nm,
the spacer width. Thus, at each tip location we collect light from several
sites. The changes in the X$^{-}$ intensity from one tip location to another
originate from a non-uniform occupation of these localization sites. To
model the localized system let us assume for simplicity that it consists of
a periodic potential in the plane, in which the electrons are randomly
distributed. In such a model the tip samples a sub-region, in which there
are $N$ sites with a probability $p$ to be occupied by electrons. The
average number of electrons is $n_{av}=pN$, and the corresponding standard
deviation is $\sigma _{n}=\sqrt{pN(1-p)}$. The average PL intensity can be
written as $I_{av}=\alpha n_{av}$, where $\alpha $ is a proportionality
factor. In this model the measured standard deviation in the X$^{-}$
intensity is given by $\sigma =\alpha \sigma _{n}$. Taking the maximum
intensity when all the sites are occupied to be $I_{max}$, one can show that 
$N=\frac{I_{max}I_{av}}{\sigma ^{2}}(1-\frac{I_{av}}{I_{max}}).$ Since $%
I_{av}$ and $\sigma $ depend on the gate voltage we can check the
consistency of this model by calculating $N$ for each gate voltage. The
inset of Fig. 4 describes $N$ as a function of the gate voltage. We can see
that indeed we get a relatively small scatter around an average value of $%
N\approx 30$. Since our tip diameter is 250 nm this implies an average
fluctuation size of $40-50$ nm. This is in a very good agreement with the
expected size, which is the spacer width (37.5 nm). It is important to note
that we have not found in the gate voltage range when the spectrum is
excitonic any area of the sample where the signal is 2DEG-like, which would
represent puddles of free electrons. This indicates that there are no large
clusters of free electrons, and the localization is of {\it single electrons}%
.

The behavior of a two-dimensional carriers gas at zero magnetic field is a
subject of a large current interest: it was recently shown that in some
material systems the conductivity undergoes a metal-insulator transition.%
\cite{Kravchenko} We wish to emphasize that the question whether the system
is initially weakly localized or metallic can not be answered by this
optical experiment. It should be noted, however, that this metal-insulator
transition has not been found in GaAs 2DEG, and it likely that the 2DEG in
our sample is initially weakly localized.

Finally, we use the fact that the location of the fluctuations is
independent of the gate voltage to image the donors distribution in the
plane. This is done by fixing the gate voltage at a convenient value and
scanning the probe to obtain a two dimensional map of the X$^{-}$ intensity.
Figure 5 shows such a map for a region of 6x6 $\mu $m$^{2}$. Regions with
large amplitudes of the X$^{-}$ intensity correspond to areas with large
donors density. We examine the statistical distribution of the X$^{-}$
intensities and find a nice Gaussian distribution. An analysis of this type
can, in principle, reveal the existence of correlation in this distribution.
In such a case a deviation from a Gaussian distribution is expected.

In conclusion, we have demonstrated the use of NSOM for imaging of charge
distribution. Our near-field PL measurements of a gated 2DEG detect the
random potential induced by the remote ionized donors, and determine the
electrons distribution in this potential. The correlation between the
appearance of the excitonic spectrum and the rise of the fluctuations
amplitude of the X$^{-}$ intensity show that the electrons are singly
localized in that potential.

This research was supported by the Minerva foundation and the Israel Academy
of Science. We thank G. Finkelstein, M. Heiblum, A. Stern and A. Yacoby for
fruitful discussions.

\figure{Fig. 1: }The evolution of the far-field PL of a 2DEG with gate
voltage (upper curves correspond to more negative gate voltage).

\figure{Fig. 2:}A. Four near field spectra taken at the same gate voltage, V$%
_{\text{g}}=-0.135$ V, and at different tip locations. B. Integrated near
field spectra of 1x1 $\mu $m$^{2}$ region (solid line) and far field
spectrum taken at 1 $\mu $m above this region (dotted line).

\figure{Fig. 3} A. The integrated charged (solid line) and neutral (doted
line) exciton intensity along an 11 $\mu $m-long line scan. The letters
denote the location of the spectra shown in Fig. 2A. B. The integrated
intensity under the X$^{-}$ peak along 11 $\mu $m-long line at three gate
voltages. The curves are normalized by dividing the value at each point by
the average of the whole line. The dashed lines are the normalized average
value for each gate voltage. The curves at $-0.135$ V and $-0.155$ V are
shifted up by $0.5$ and $1.0$, respectively, for clarity (where $1.0$ is
clearly the normalized average value).

\figure{Fig. 4: }The standard deviation of the relative fluctuations
amplitude of the X$^{-}$ intensity along an 11 $\mu $m-long line, as a
function of gate voltage. The averaged spectra at V$_{\text{g}}$ $=-0.35$ V
and V$_{\text{g}}=-0.41$ V are also shown. Inset: The number of sites under
the tip, calculated for various gate voltages. The calculation method is
explained in the text.

\figure{Fig. 5: }Two-dimensional image of the X$^{-}$ integrated intensity
for a $6\times 6$ $\mu $m$^{2}$ region.

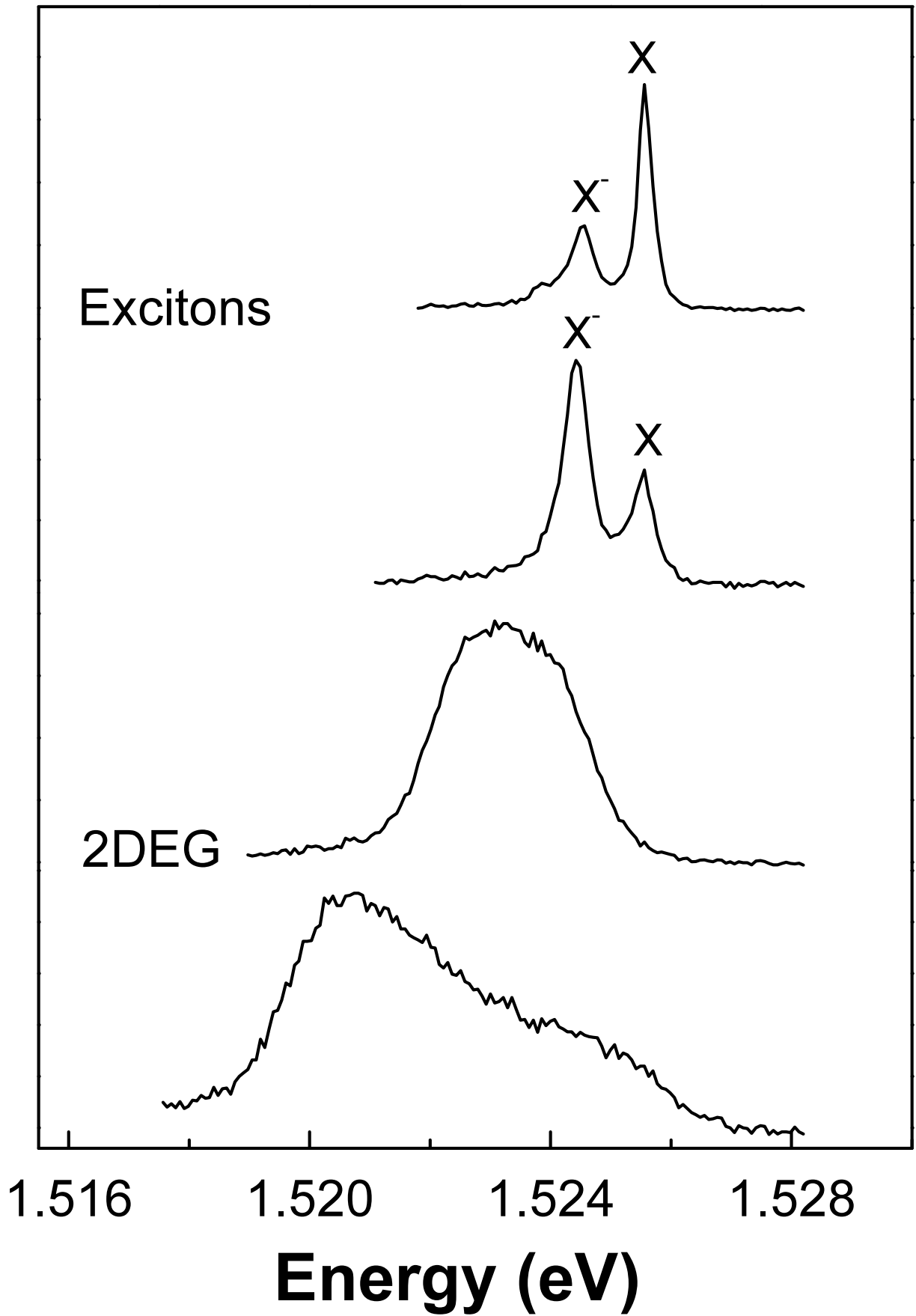
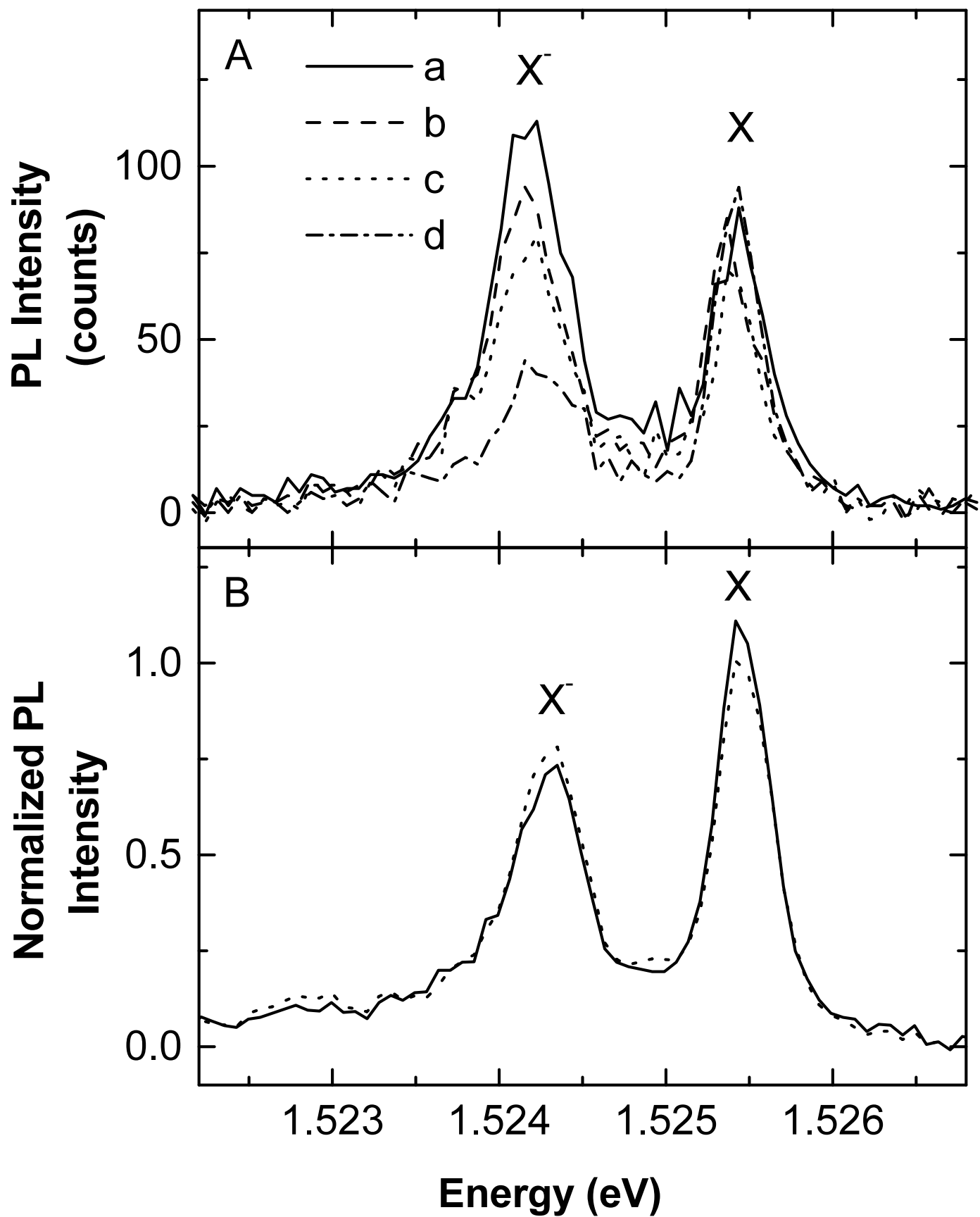
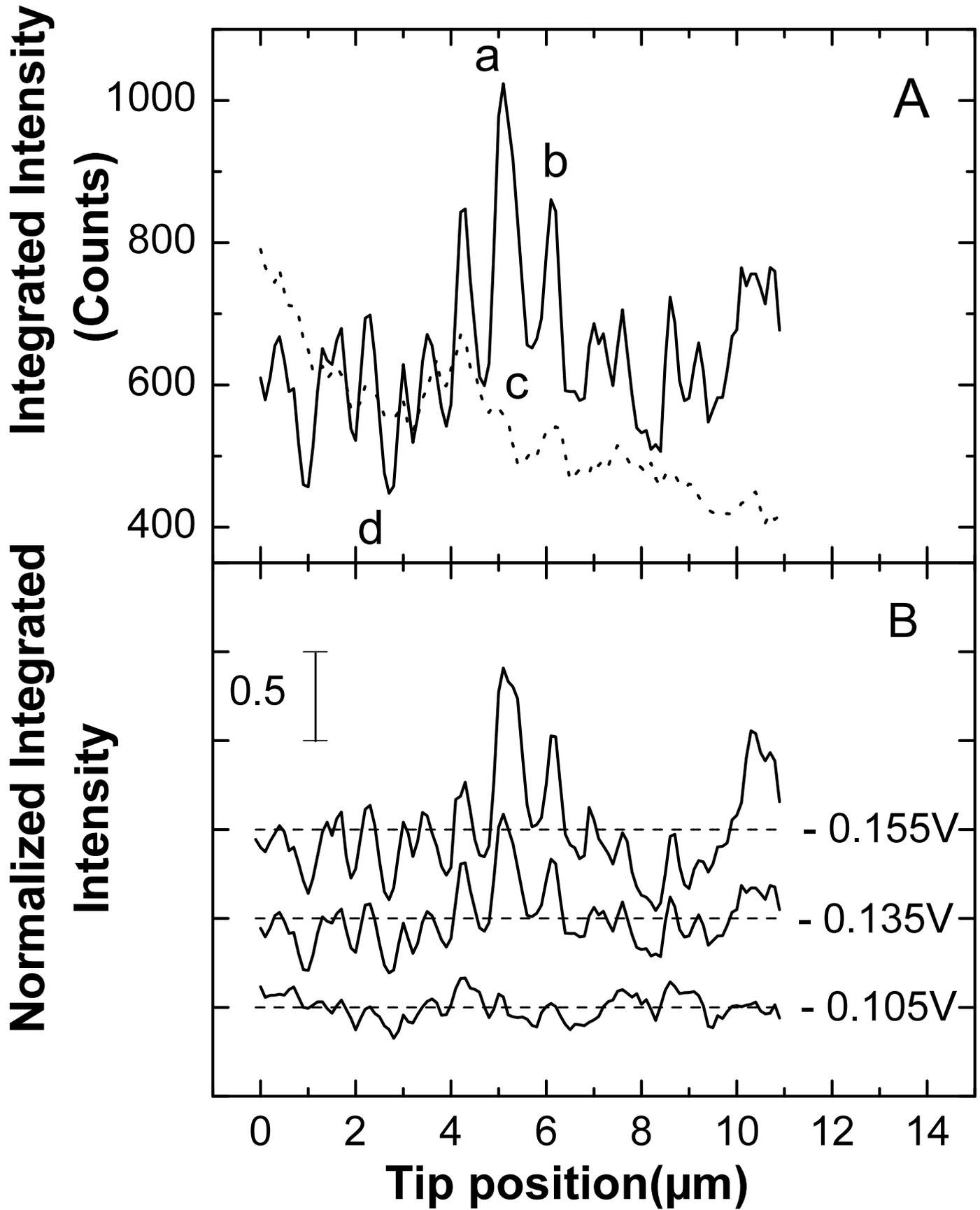
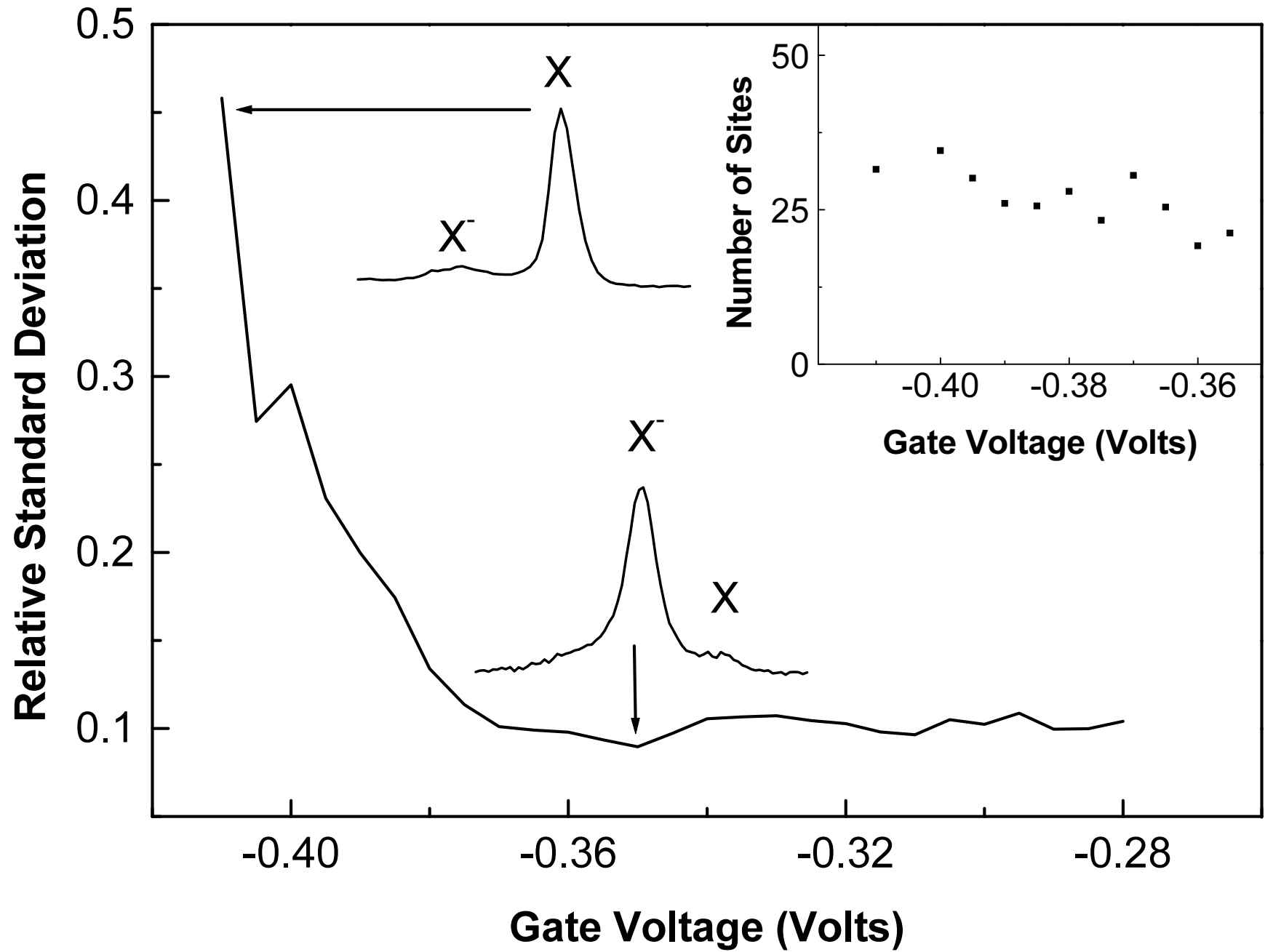
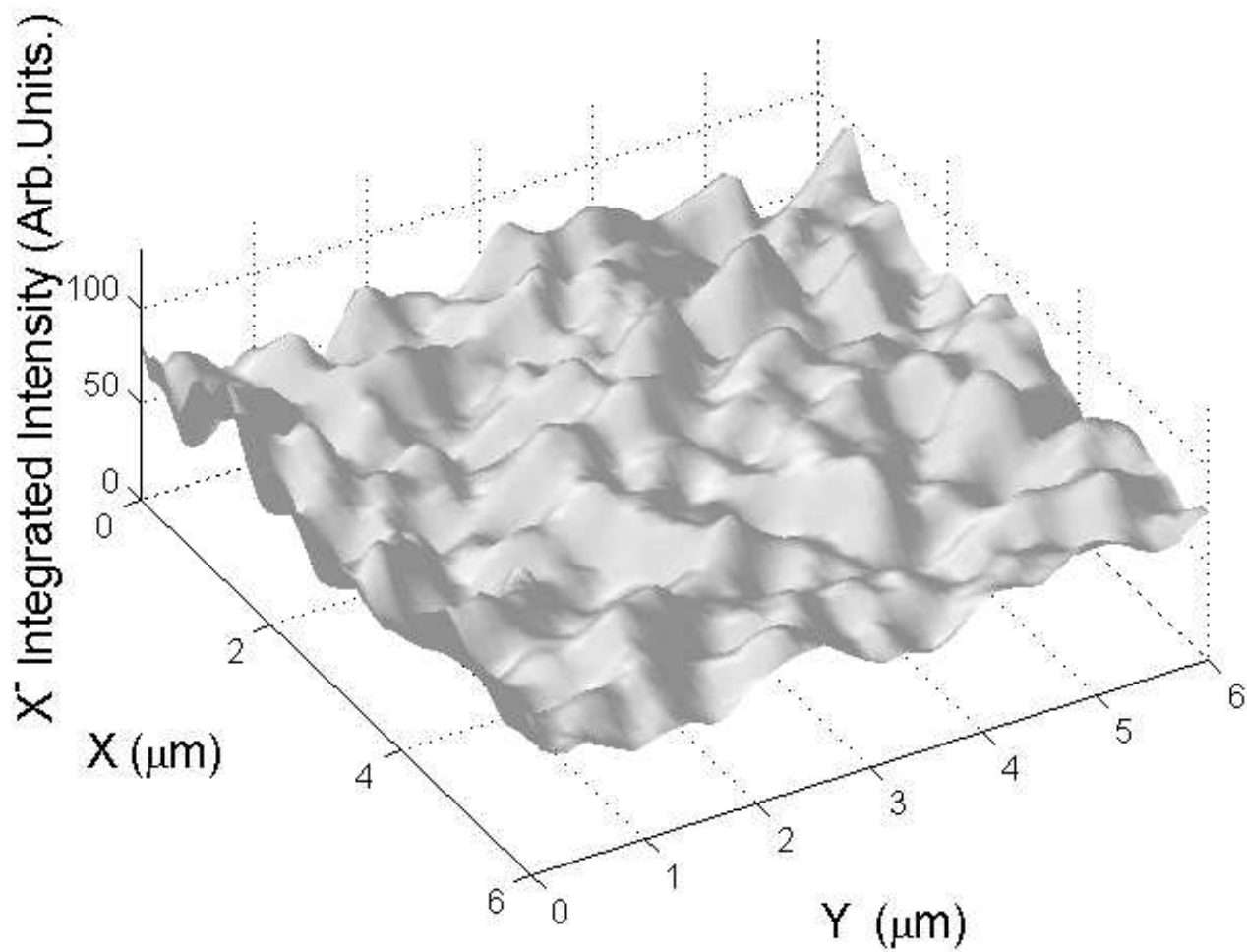
\end{document}